\documentclass[twocolumn]{aastex701}
\usepackage{amsmath}

\received{December 12, 2025}
\revised{January 17, 2026}
\accepted{January 19, 2026}

\newcommand{\rv}{$R_V$} 
\newcommand{\rnot}{$R_0$} 
\newcommand{\gaia}{\textit{Gaia}}

\begin{document}

\title{The Effect of a Non-universal Extinction Curve on the Wesenheit Function and Cepheid Distances}

\author[orcid=0000-0001-9439-604X,gname='Dorota',sname='Skowron']{Dorota M. Skowron}
\affiliation{Astronomical Observatory, University of Warsaw, Aleje Ujazdowskie 4, 00-478 Warsaw, Poland}
\email[show]{dszczyg@astrouw.edu.pl}  

\author[orcid=0000-0001-9256-5516,gname=Morgan, sname=Fouesneau]{Morgan Fouesneau} 
\affiliation{Max Planck Institute for Astronomy, K\"{ o}nigstuhl 17, 69117 Heidelberg, Germany}
\email{fouesneau@mpia.de}

\author[orcid=0000-0000-0000-0000,gname=Ronald, sname=Drimmel]{Ronald Drimmel} 
\affiliation{INAF - Osservatorio Astrofisico di Torino, via Osservatorio 20, 10025 Pino Torinese (TO), Italy}
\email{ronald.drimmel@inaf.it}

\author[orcid=0000-0000-0000-0000,gname=Shourya, sname=Khanna]{Shourya Khanna} 
\affiliation{INAF - Osservatorio Astrofisico di Torino, via Osservatorio 20, 10025 Pino Torinese (TO), Italy}
\email{shourya.khanna@inaf.it}

\begin{abstract}

The Wesenheit function is widely used to reduce the effects of interstellar reddening in distance measurements. Its construction, however, relies on the assumption of a universal extinction curve and on fixed values of the total-to-selective extinction ratio, \rv. Recent studies have shown that \rv\ varies significantly across the Milky Way and between different galaxies, raising concerns about systematic biases in Wesenheit magnitudes and period–Wesenheit relations. In this work, we discuss the impact of non-universal extinction on Wesenheit indices by combining the \rv-dependent extinction curve with a grid of stellar atmosphere models. We compute the integrated extinction in optical and near-infrared passbands, derive \rv-dependent $R$ coefficients for multiple Wesenheit indices, and examine how changes in \rv\ propagate into Wesenheit magnitudes and Cepheid distances in our Galaxy.
We find that the $R$ coefficients in the Wesenheit functions vary strongly with \rv. For classical Cepheids in the Milky Way disk, variations of \rv\ within the typical observed range ($2.6-3.6$) can lead to substantial differences in the Wesenheit function, reaching $\pm0.7$ mag from the mean for the Gaia-based Wesenheit index $W_G$ and resulting in distance errors of almost $40$\%. Near-infrared Wesenheit indices are much less sensitive to \rv\ changes.
Our results clearly show that accounting for variable \rv\ is essential when applying period--Wesenheit relations, particularly in the optical regime, or that near or mid infrared based distances should be used.
While we present this effect for classical Cepheids, it applies to all pulsating stars for which period--Wesenheit relations are used to infer distances.
\end{abstract}


\section{Introduction}

The Wesenheit\footnote{Wesenheit is a German word for "intrinsic essence"} function, $W$, is an expression for the magnitude of a star that is free of interstellar extinction and reddening. It was conceptually introduced by \cite{vandenBergh1975}, although the idea of constructing a reddening-free magnitude traces back to \cite{Johnson1953}. The term “Wesenheit” itself was later introduced by \cite{Madore1982}, who also explored its detailed properties and applications, using it to calibrate the period--luminosity relation for classical Cepheids. A comprehensive historical overview of the development and evolution of the Wesenheit function, as well as its underlying physical basis can be found in \cite{Madore2024}.

While the Wesenheit function can be applied to any type of star, it has been used primarily as a luminosity indicator in the period–luminosity (PL) relation for pulsating stars, such as Cepheids, RR~Lyrae, or long-period variables. This relation, expressed as $W = a \log P + b$, is commonly referred to as the period–Wesenheit (PW) relation. Its use significantly reduces the scatter relative to the standard PL relation by effectively accounting for interstellar extinction \citep[see Fig.~6 in][]{Soszynski2008}.
This is particularly important in the case of classical Cepheids, which serve as standard candles for measuring precise distances within the Milky Way and in galaxies well beyond the Local Group, making them excellent tracers of young stellar populations and key contributors to the calibration of the cosmic distance scale and the determination of the Hubble constant \citep{Casertano2025}.
Therefore, the subsequent analysis will be focused on classical Cepheid variable stars.

The main assumption in constructing the Wesenheit function is that the ratio of total-to-selective extinction, $R$, is {\em universal within a galaxy and known a priori} (i.e., a universal extinction curve). Otherwise, if $R$ is not correct or not universal, the Wesenheit magnitude will not be reddening-free. 
In the optical, the most commonly used total-to-selective extinction ratio is
\begin{equation}
    R_V = \frac{A_V}{E(B-V)} = \frac{A_V}{A_B - A_V}\, ,
    \label{eqn:rv}
\end{equation}
which is assumed to have a constant mean value of around $3.1$ in the Milky Way \citep[e.g.,][]{Schultz1975, Fitzpatrick1999}. However, growing evidence indicates that \rv\ is not constant within the Galaxy, but it varies significantly, particularly in regions of high extinction \citep{Fitzpatrick2007,  Schlafly2016, Nataf2016, Maiz-Apellaniz2018, Fouesneau2022, Zhang2023, Zhang2025}. This is especially important for classical Cepheids, which belong to the young population and are therefore confined to the Galactic disk, where the extinction is highest.
The adverse consequences of a variable \rv\ on Cepheid distance estimates for Galactic structure studies have already been reported by \cite{Skowron2025}, and this letter aims to explore this effect in greater detail.

In the following section, we describe the Wesenheit function and how it is defined for a given set of passbands, including the $R$ coefficient.  Section 3 details how we apply the extinction curve to derive extinctions for a variety of passbands for different values of \rv. Section 4 investigates the effect of a variable \rv\ on the Wessenheit magnitudes and distances, and we summarize our conclusions in Section 5.

\section{The Wesenheit Function Definitions}

The Wesenheit function (also called the Wesenheit index or the Wesenheit magnitude) is traditionally calculated from apparent magnitudes at two different passbands characterised by wavelengths $\lambda_1$ and $\lambda_2$, as:
\begin{equation}
    W_{\lambda_1,\lambda_2} = m_{\lambda_2} - R_{\lambda_1,\lambda_2} \times (m_{\lambda_1} - m_{\lambda_2})
    \label{eqn:wesenheit}
\end{equation}
where $R_{\lambda_1,\lambda_2}$ is the ratio of total-to-selective extinction:
\begin{equation}
    R_{\lambda_1,\lambda_2} = \frac{A_{\lambda_2}}{E(m_{\lambda_1} - m_{\lambda_2})} = \frac{A_{\lambda_2}}{A_{\lambda_1} - A_{\lambda_2}}
    \label{eqn:r2}
\end{equation}
with $A_{\lambda}$ denoting the extinction and $E(m_{\lambda_1} - m_{\lambda_2})$ representing the color excess.
Another representation of the Wesenheit magnitude is composed of three passbands:
\begin{equation}
    W_{\lambda_1,\lambda_2,\lambda_3} = m_{\lambda_1} - R_{\lambda_1,\lambda_2,\lambda_3} \times (m_{\lambda_2} - m_{\lambda_3})
\end{equation}
where:
\begin{equation}
    R_{\lambda_1,\lambda_2,\lambda_3} = \frac{A_{\lambda_1}}{E(m_{\lambda_2} - m_{\lambda_3})} = \frac{A_{\lambda_1}}{A_{\lambda_2} - A_{\lambda_3}}
    \label{eqn:r3}
\end{equation}

\begin{deluxetable*}{llll}[htb]
\tablewidth{0pt}
\tablecaption{Wesenheit index definitions used in various studies and calculated in this study using the extinction curve of \citet{Gordon2023}.\label{tab:wesenheit}
}
\tablehead{
\colhead{Wesenheit index} & \colhead{$R$ definition} & \colhead{$R$ value} & \colhead{$R$ origin} }
\startdata 
$W_{VI}=I-R_{VI}(V-I)$ & $R_{VI}=A_I/E(V-I)$ & $1.55$  & $R_V=3.23$, \cite{Cardelli1989} \\
                       &                     & $1.387$ & $R_V=3.1$, \cite{Fitzpatrick1999} \\
                       &                     & $1.335$ & $R_V=3.1$, \cite{Gordon2023}\\
\hline
$W_{JK}=K-R_{JK}(J-K)$ & $R_{JK}=A_K/E(J-K)$ & $0.679$ & $R_V=3.1$, \cite{Cardelli1989}\\
                       &                     & $0.688$ & $R_V=3.23$, \cite{Cardelli1989}\\
                       &                     & $0.553$ & $R_V=3.1$, \cite{Gordon2023}\\
\hline
$W_{VK}=K-R_{VK}(V-K)$ & $R_{VK}=A_K/E(V-K)$ & $0.129$ & $R_V=3.23$, \cite{Cardelli1989}\\
                       &                     & $0.090$ & $R_V=3.1$, \cite{Gordon2023}\\
\hline
$W_{H}=H-R_{H}(V-I)$   & $R_{H}=A_H/E(V-I)$  & $0.365$ & $R_V=3.1$, \cite{Cardelli1989} \\
                       &                     & $0.35$  & $R_V=3.1$, \cite{Fitzpatrick1999} \\
                       &                     & $0.321$ & $R_V=3.1$, \cite{Gordon2023}\\
\hline
$W_G=G-R_G(G_{BP}-G_{RP})$ & $R_G=A_G/E(G_{BP}-G_{RP})$ & $1.9$ & empirical, \cite{Ripepi2019} \\ 
                           &                            & $1.67$ & $R_V=3.1$, \cite{Gordon2023}\\
\enddata 
\end{deluxetable*}

In practice, the choice of which Wesenheit index to use is dictated by the passbands for which observational data are available. Traditionally, $B$ and $V$ bands have been employed; however, following the publication of catalogs containing over $9,500$ Cepheids in the Magellanic Clouds by the Optical Gravitational Lensing Experiment (OGLE), the $V$ and $I$ bands have become widely adopted \citep{Soszynski2008, Soszynski2015}, while additional observations from near-IR surveys 
have led to the use of a variety of Wesenheit indices that combine the $V$, $J$, $H$, and $K_S$ magnitudes. 
The use of classical Cepheids as standard candles, where distance measurements are based on Hubble Space Telescope (HST) observations in the optical ($V=F555W$, $I=F814W$) and near-infrared ($H=F160W$) bands, relies on the near-IR Wesenheit index, $W_H$, constructed from these three HST passbands \citep{Riess2009}.
With the availability of all-sky data for classical Cepheids from the Gaia mission, the Wesenheit magnitude $W_G$ was introduced, with the three-band variant exhibiting the lowest scatter \citep{Ripepi2019}.

In the following analysis, we focus on the most often used Wesenheit functions, listed in Table~\ref{tab:wesenheit}, although numerous other combinations have also been studied \citep[e.g.][]{Inno2016, Breuval2022, Bhardwaj2024}.

\section{Extinction Curve}

To calculate the $R$ coefficient in different bands, we need to assume a relation between interstellar dust extinction and wavelength. Commonly used \rv\ dependent extinction curves in the Milky Way include those of \cite{Cardelli1989} and \cite{Fitzpatrick1999}, and more recently \cite{Gordon2009, Fitzpatrick2019, Gordon2021, Decleir2022, Zhang2025}, which cover various wavelength ranges.  However, because these studies used different samples, methods, and wavelength ranges, combining the resulting extinction curves to cover the full ultraviolet-mid-infrared range could introduce artificial discontinuities between wavelength regions.
Therefore, for the following analysis, we adopt the extinction curve of \citet[][hereafter G23]{Gordon2023}, which is the first \rv\ dependent extinction relation valid over the full wavelength range.

G23 define an extinction curve normalized to the monochromatic extinction at 550\,nm.
However, we stress that the wavelength dependence of their extinction curve in the optical range does not differ significantly from past formulations  (e.g., \citealt{Fitzpatrick1999, Fitzpatrick2019}). Nevertheless, their normalization to a monochromatic extinction leads to differences that we highlight below. 

To avoid confusing notations and to make a clear distinction between measured quantities and model parameters, we will henceforth denote $A_0$ and $R_0$ for the extinction and total-to-selective extinction ratio parameters of the G23 extinction curve, respectively, and reserve the notation $A_X$, $R_X$ for integrated effects within a given passband $X$ (or monochromatic measurements).

For a star of spectral energy distribution $f_\lambda$ and the G23 extinction curve $A_\lambda (A_{0}, R_{0})$, we define the attenuation in the passband $X$ of throughput $T_X$ as
\begin{equation}
\label{eq-Aband-def}
A_{X} = -2.5 \log_{10} \frac{\int \gamma_X(\lambda) T_X(\lambda) f_\lambda e^{-A_\lambda} d\lambda}{\int \gamma_X(\lambda) T_X(\lambda) f_\lambda d\lambda},
\end{equation}
with $\gamma_X(\lambda) = 1$ for energy-counting (e.g., Johnson $B$, $V$, $I$ bands), or $\gamma_X(\lambda) = \lambda / (hc)$ for photon-counting (e.g, Gaia $G$, $G_{BP}$, $G_{RP}$; 2MASS $J$, $K{_s}$) photometric system.
As $A_{X}$ is not the weighted average monochromatic extinction through $T_X$, we must calculate the integrated effect for any spectrum of interest.

We used the Atlas-9 (2004; \citealt{Castelli2003}) atmosphere library to compute a grid of
$17,603$ stellar spectra references spanning effective temperatures $ T_{\rm eff}\in[2,500; 8,000]$\,K, surface gravities $\log g \in [-0.5; 5.5]$\,dex, and metallicities $[M/H] \in [-5; +1]$\,dex.
We expand this collection by applying the G23 extinction curve on a grid of $A_{0} \in [0, 10.0]$ by step of $0.5$ mag and $R_{0} \in [2.3, 2.6, 3.1, 3.6, 4.1, 4.6, 5.1, 5.6]$. Finally, we use Pyphot \citep{Fouesneau_pyphot_2025} to compute the photometric calculations in the relevant passbands for the present work, namely: Johnson $B$ \& $V$; Cousins $I$; Gaia DR3 $G$, $G_{BP}$, $G_{RP}$ and 2MASS $J$, $H$, $K_S$.\footnote{We took the transmission curves from the SVO-profile service: \url{http://svo2.cab.inta-csic.es/theory/fps/}; \citet{svofps2020}}

\begin{figure} 
\includegraphics[width=\linewidth]{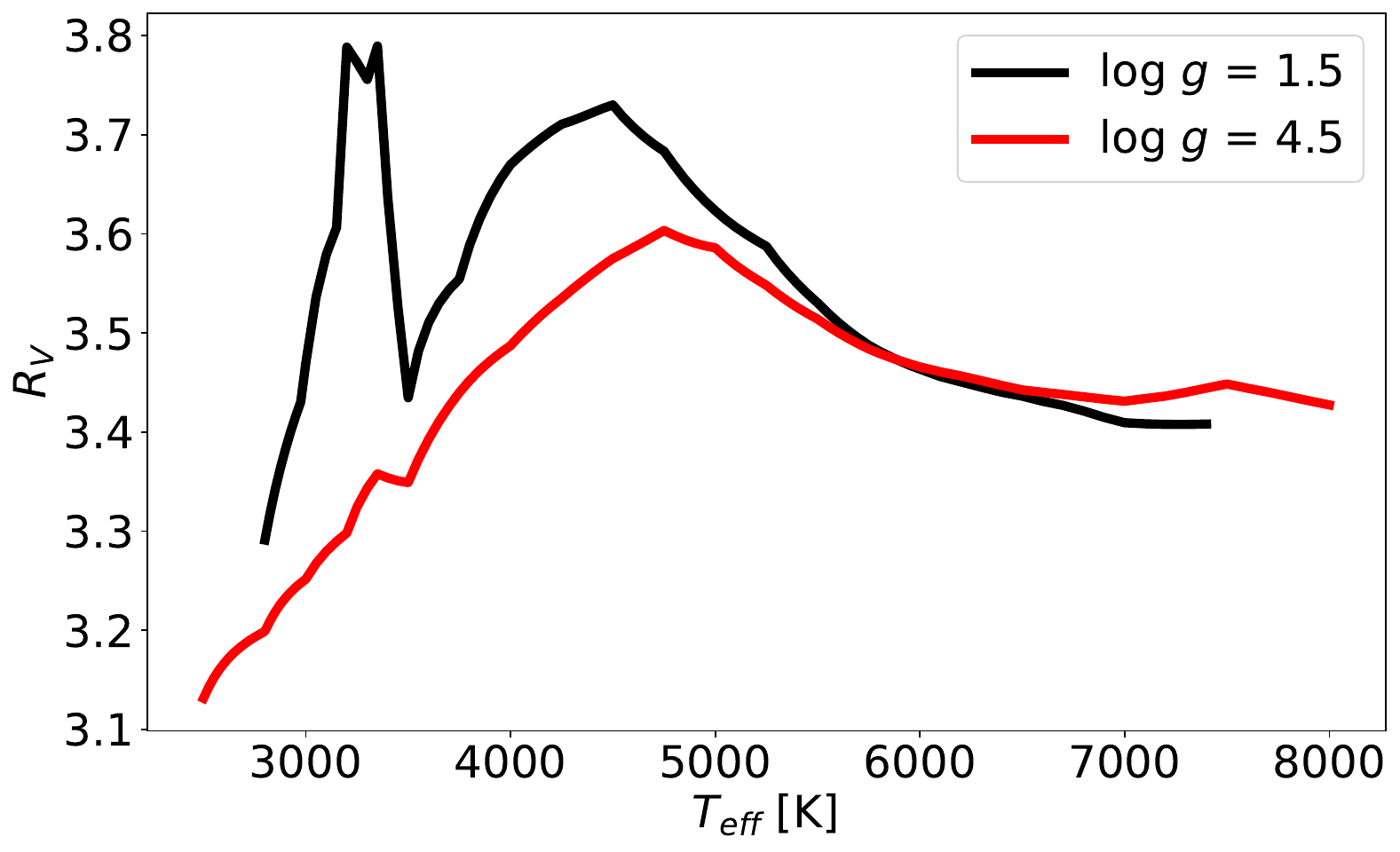} 
\caption{\rv\ as a function of effective temperature $T_{\rm eff}$ for giants ($\log g = 1.5$) and main sequence stars ($\log g = 4.5$), assuming a G23 extinction curve with \rnot$= 3.1$. 
\label{fig:rvtemp}} 
\end{figure}

It is important to note that the $R_{0}$ parameter that controls the slope of the extinction curve in G23\footnote{Referred to as "$R(V)$" in their paper, and as "\texttt{Rv}" in their python package.} is not equivalent to \rv.
In the past, this extinction parameter was often ``calibrated'' using actual photometric \rv\ measurements, so that \rnot\ was approximately equal to \rv. However, in the case of G23, the parameter \rnot\ is calibrated on spectroscopic measurements of $R_{55} = A(550)/(A(440)-A(550))$, where $A(550)$ and $A(440)$ are the monochromatic extinctions at 440 and 550~nm, leading to a substantial offset with respect to \rv. To illustrate this offset and the dependence of \rv\ on the source spectra, Figure~\ref{fig:rvtemp} shows how \rv\ varies with respect to the effective temperature for stars and different $\log g$, for a fixed extinction curve with $R_0 = 3.1$. We note that at the typical temperature of Cepheids ($T_{eff} = 5500$\,K), and for hotter stars, the variation in \rv\ is minor, while the variation in \rv\ for stars cooler than about 5000\,K, which includes giants and lower main sequence stars, can be quite large.

\begin{figure} 
\includegraphics[width=\linewidth]{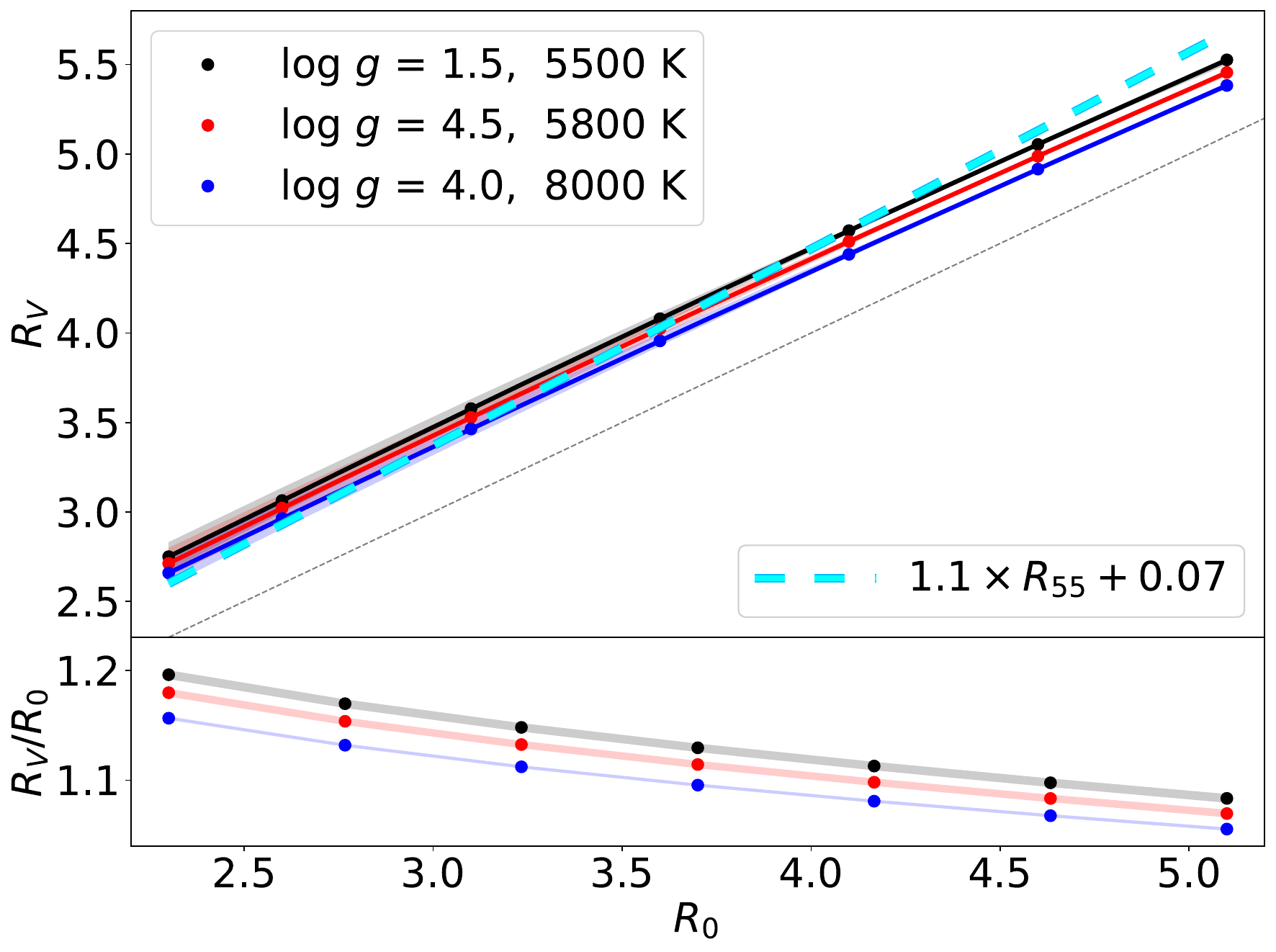} 
\caption{\rv\ as a function of \rnot, shown for three different types of stars. In the upper panel, the shaded region captures the spread over 0.5 $<A_{0}<$ 9.5, while the points are the median values. The black solid curve is the best-fit polynomial for Cepheids (Eq.~\ref{eqn:r0torv}), while the cyan line is the proposed relation between \rv\ and $R_{55}$ of \citet{Zhang2025}. The grey dotted line indicates the identity relation. The lower panel shows the ratio \rv/\rnot\ with the spread capturing the uncertainty in the polynomial fit. 
\label{fig:rvr0}} 
\end{figure}

In Figure~\ref{fig:rvr0} (upper panel), we show the relation between \rv\ and \rnot\ for three different stellar types, including a typical Cepheid ($T_{\rm eff} = 5500\,{\rm K}, \; \log g = 1.5$). In each case, we show the relation for $0.5 < A_{0} < 9.5$\,mag with the spread indicated by the shaded regions, and the median indicated by points. There is an apparent offset between \rv\ and \rnot\, which we characterize  by fitting a second-order polynomial to the median points, finding:
\begin{equation}
    R_V = 0.2267 + 1.145 \, R_0 - 0.021 \, R_0^2 \, 
    \label{eqn:r0torv}
\end{equation} 
where the coefficients have uncertainties of no more than 2\%, and they remain unchanged within $T_{\rm eff}$ and $\log g$ range typical for Cepheids.
While this offset remains nearly constant as \rnot\ is varied, the lower panel of Figure~\ref{fig:rvr0} shows the ratio \rv/\rnot\ to highlight the subtle variation with respect to \rnot, where the shaded region illustrates the uncertainty in the ratio from our polynomial fit.
Because the parameter \rnot\ is calibrated on measurements of $R_{55}$, we found that the offset of \rnot\ with respect to \rv\ (Eq.~\ref{eqn:r0torv}) is quite similar to the offset of $R_{55}$ with respect to \rv, as specified by \citet{Zhang2025}, i.e. $R_V = 1.1 \, R_{55} + 0.07$.  

\section{Discussion}

\begin{figure*}
\begin{center}
\begin{tabular}{ccc}
\includegraphics[width=0.32\linewidth]{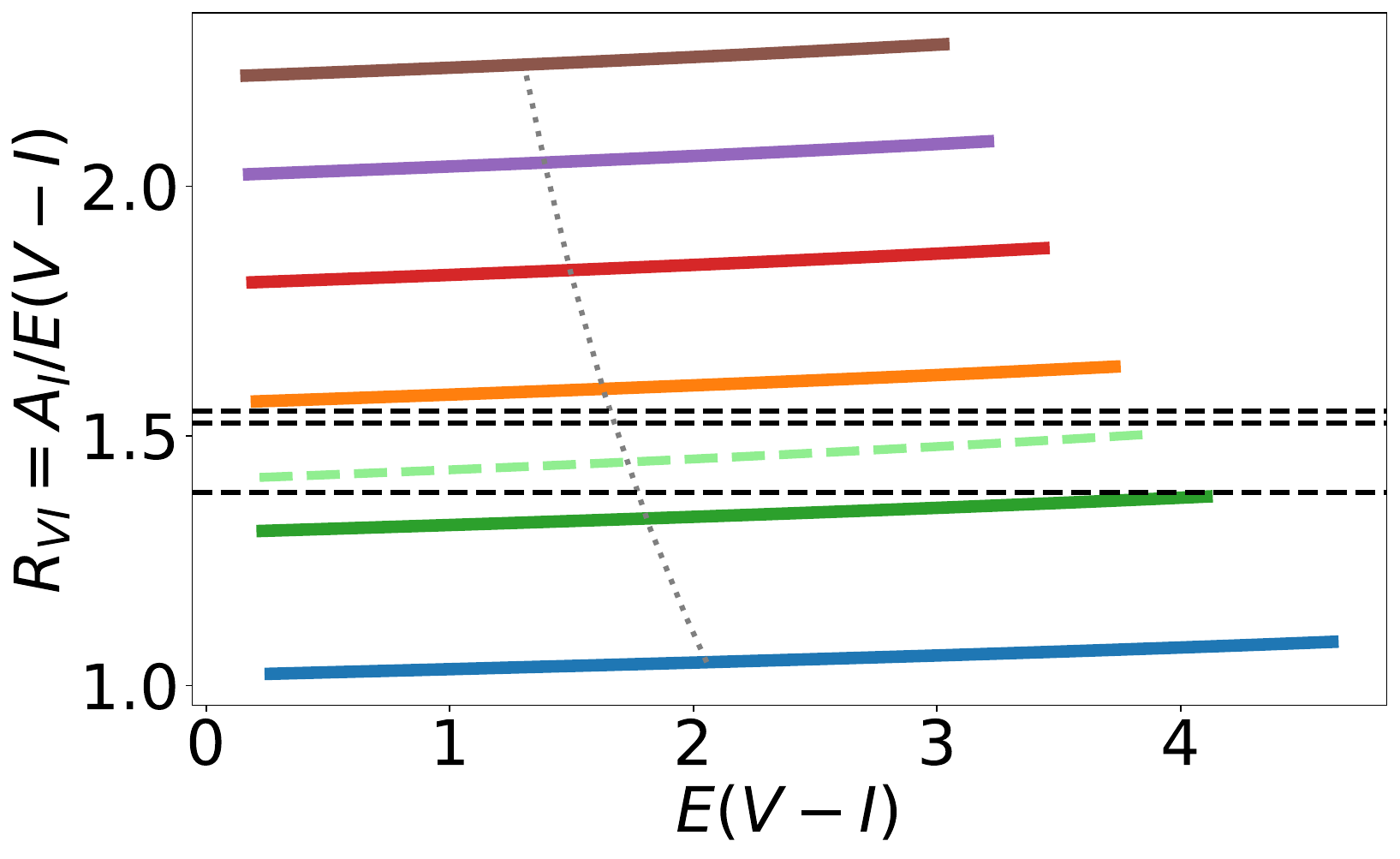} &
\includegraphics[width=0.32\linewidth]{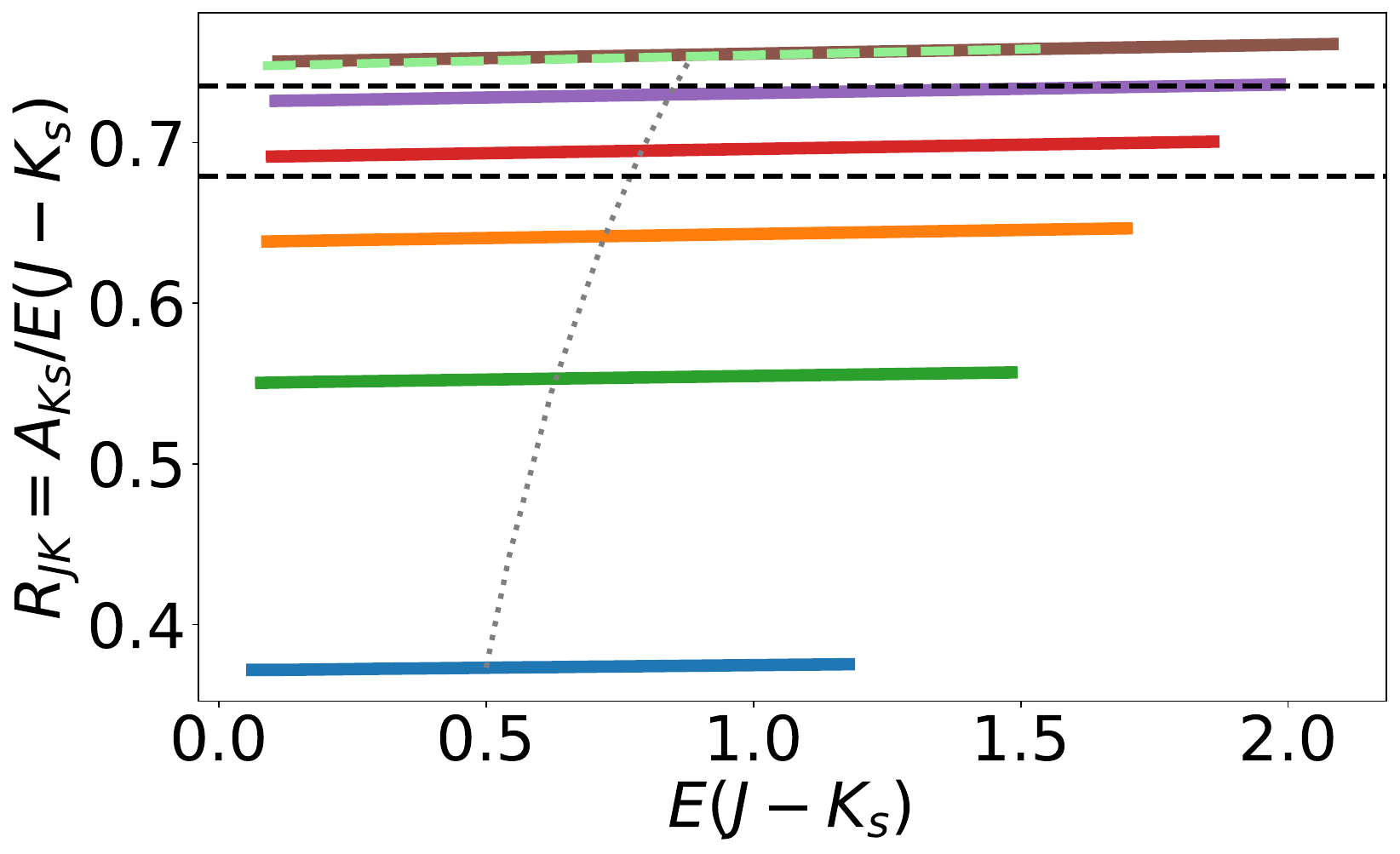} &
\includegraphics[width=0.32\linewidth]{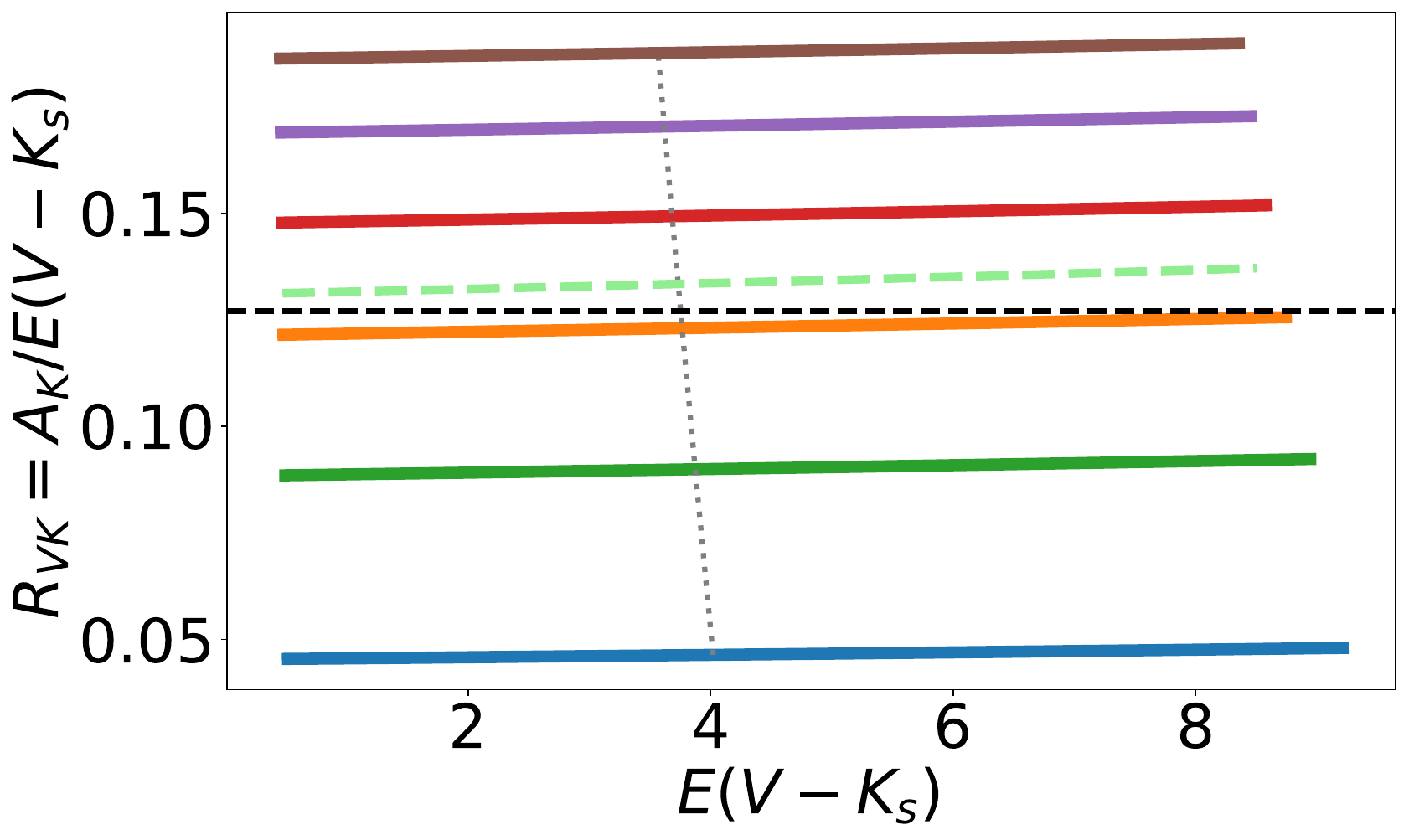} \\
\includegraphics[width=0.32\linewidth]{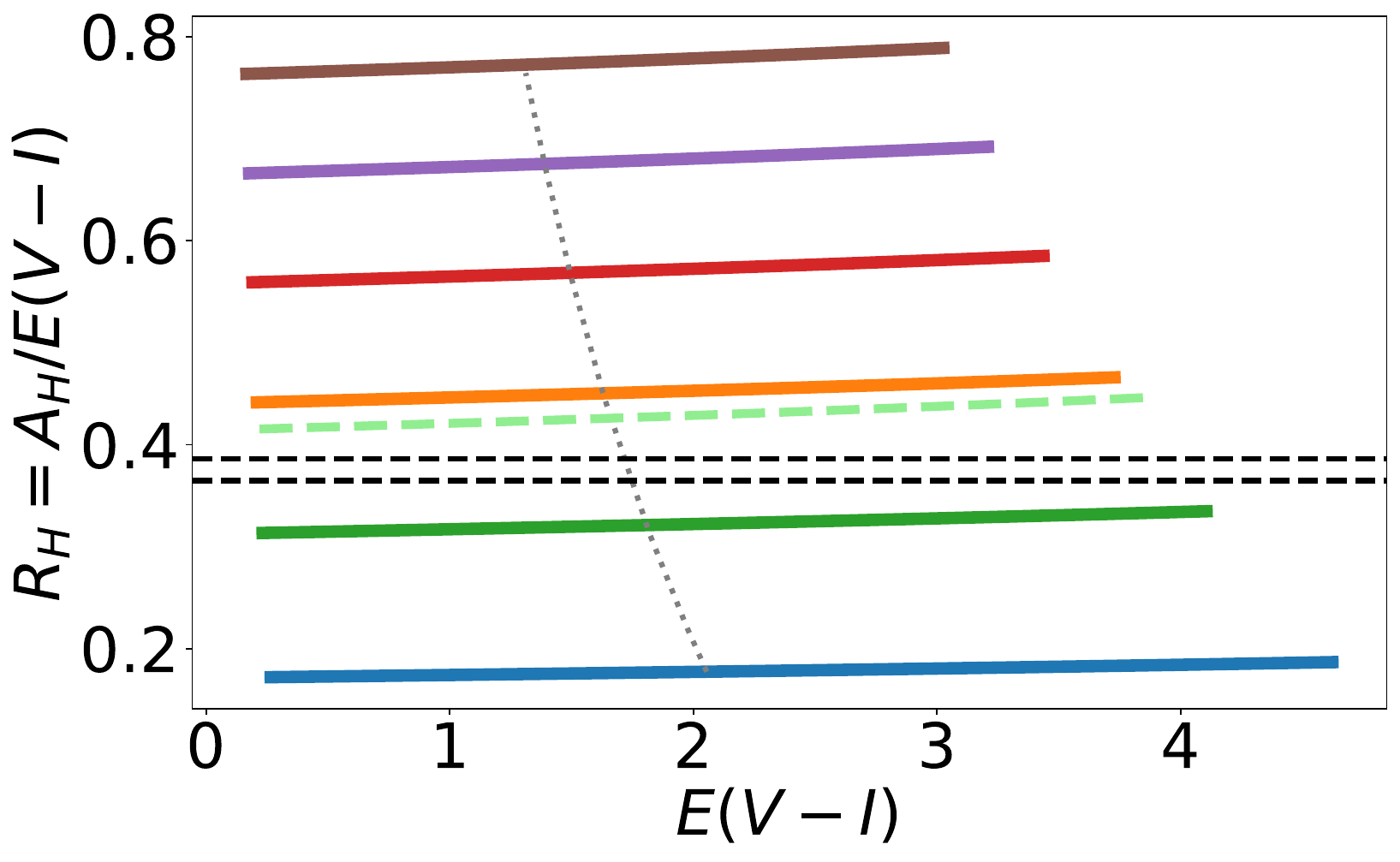} &
\includegraphics[width=0.32\linewidth]{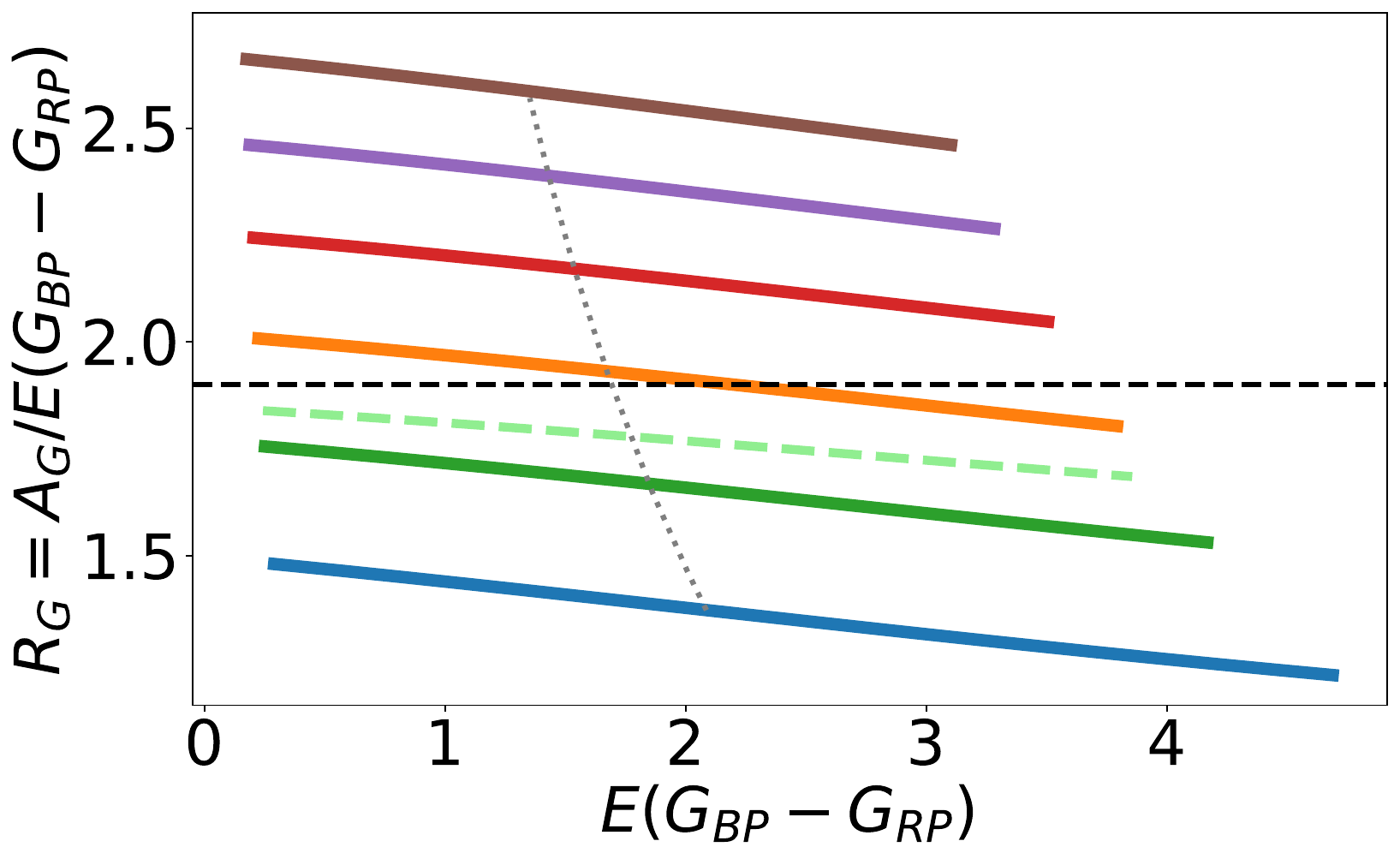} & 
\includegraphics[width=0.32\linewidth]{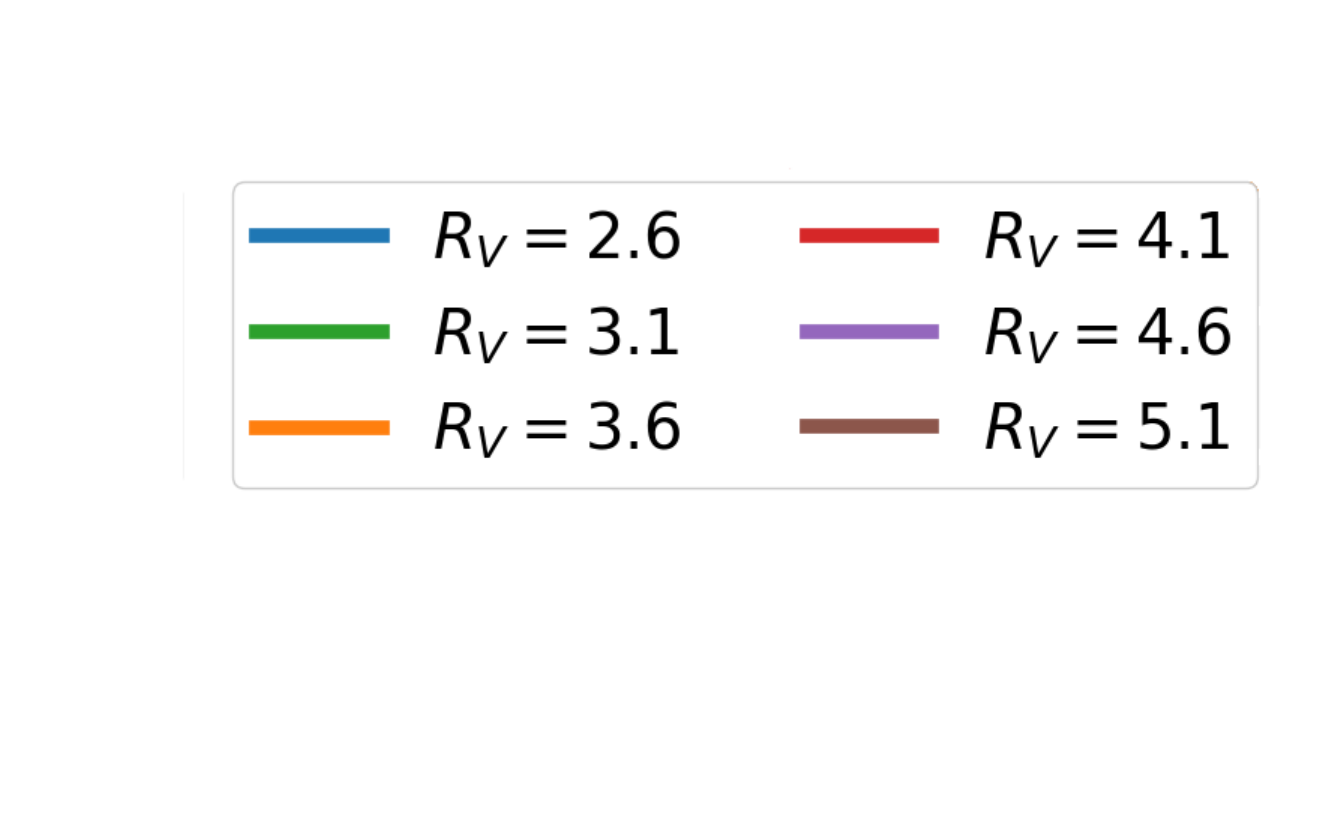} \\
\end{tabular}
\end{center}
\caption{The dependence of the $R$ coefficient, used in the definition of a given Wesenheit index, on $R_V$, as a function of the color excess $E$ for various passbands, calculated using a typical Cepheid source spectra. Colors mark different \rv\ values, according to the legend. Solid lines assume the G23 extinction curve, while the light-green dashed line is the \cite{Fitzpatrick1999} extinction curve, but only for $R_V=3.1$. Black dashed lines show constant $R$ coefficients adopted in various studies, and the grey dotted line represents reddening values typical for Milky Way Cepheids.\label{fig:Rcoefficients}}
\end{figure*} 

Recently \cite{Zhang2025} measured the \rv\ parameter within the Milky Way and the Magellanic Clouds using low-resolution optical spectra from \gaia\ for $\sim130$ million stars. The resulting three-dimensional maps provide \rv\ values as a function of distance, enabling more accurate extinction corrections than those obtained by assuming a constant \rv\ value.
The mean differential \rv\ value in the Galaxy is $3.11$ with a $1\sigma$ range from $2.60$ to $3.88$, and the average integrated \rv\ is $3.08$ with a $1\sigma$ range from $2.86$ to $3.28$, although values as low as $2.00$ and as high as $8.00$ are observed. Both average estimates are consistent with $R_V=3.1$, as found in many previous studies of the Milky Way.
This section shows how varying extinction curve, as parameterized by \rv, affects the Wesenheit magnitudes and distances based on PW relations. That is, we fix the extinction curve using a specific value of the parameter \rnot, but in this section translate these \rnot\ values to the more commonly used \rv\ parameter used in the past to parameterize extinction curves, using Eq.~\ref{eqn:r0torv} specific to Cepheids.

\subsection{The impact of \rv\ on the $R$ coefficient}

The Wesenheit index parameter $R$ quantifies the ratio of total to selective extinction for a given combination of passbands (Eqns.~\ref{eqn:r2},\ref{eqn:r3}) and therefore varies across specific definitions of the Wesenheit index. Figure~\ref{fig:Rcoefficients} shows the $R$ coefficient as a function of the color excess $E$, where each panel corresponds to a different definition of $W$ and $R$, listed in Table~\ref{tab:wesenheit}. The color coding in each panel represents distinct \rv\ values, and the grey dotted line marks the reddening values typical for Milky Way Cepheids. 
Light-green dashed lines represent $R$, when instead of G23, the \cite{Fitzpatrick1999} extinction curve for $R_V = 3.1$ is used, and black dashed lines show constant values used in the literature.

\begin{figure} 
\vspace{5pt}
\includegraphics[width=\linewidth]{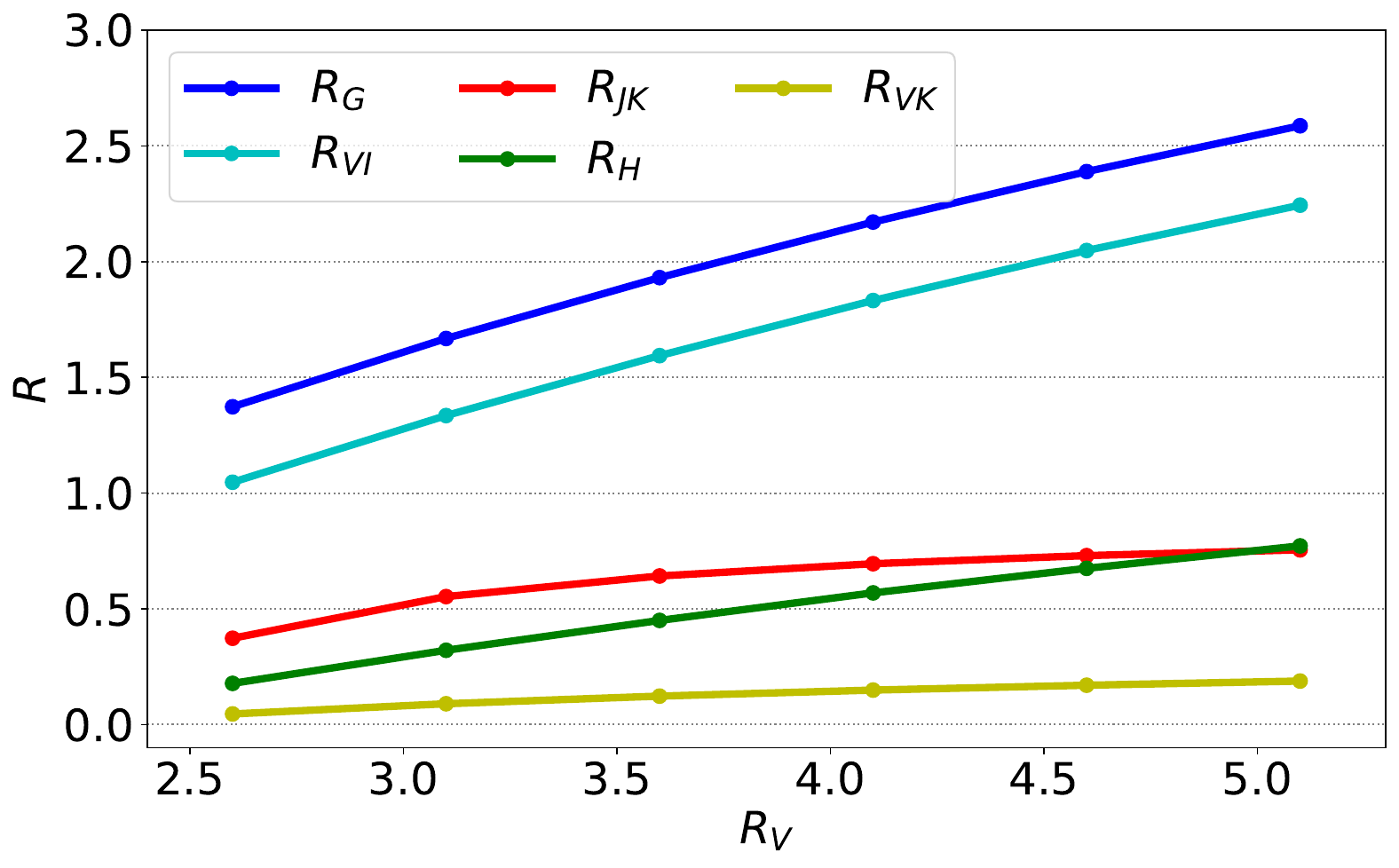} 
\caption{The $R$ parameter used in the definition of various Wesenheit indices as a function of \rv. Each color represents a different Wesenheit index parameter $R$, defined in Table~\ref{tab:wesenheit}, and plotted in Figure~\ref{fig:Rcoefficients}.
\label{fig:r_vs_rv}} 
\end{figure}

First, it is worth noting that for most definitions of $R$, the lines corresponding to different \rv\ values are not horizontal, i.e., they depend on the reddening. This indicates that, despite common understanding, $R$ is not constant over reddening for a fixed extinction curve, as reddening is not perfectly proportional to extinction. This effect is most pronounced for the Gaia coefficient $R_G$ where the colored lines exhibit the highest slope, due to the large width of the Gaia passbands.

However, a more important conclusion from Figure \ref{fig:Rcoefficients} is that the value of $R$ varies significantly with \rv.  For example, in the first panel of this figure, we see that for a typical Cepheid reddening (indicated by the gray dotted line), the value of $R_{VI}$ changes from $1.05$ to $1.6$ for \rv\ varying from 2.6 to 3.6, which is a typical \rv\ range in the Milky Way. In Section~\ref{sec:magdist} we will show how this impacts the Wesenheit magnitude and the resulting distance determinations.

In Figure~\ref{fig:r_vs_rv}, we summarize plots from Figure~\ref{fig:Rcoefficients}, by showing how the $R$ coefficient changes with \rv\ for different Wesenheit functions. Each line corresponds to one panel from Figure~\ref{fig:Rcoefficients}, assuming a mean Cepheid reddening as indicated by the grey dotted lines.

\begin{figure*}[bht]
\includegraphics[width=0.49\linewidth]{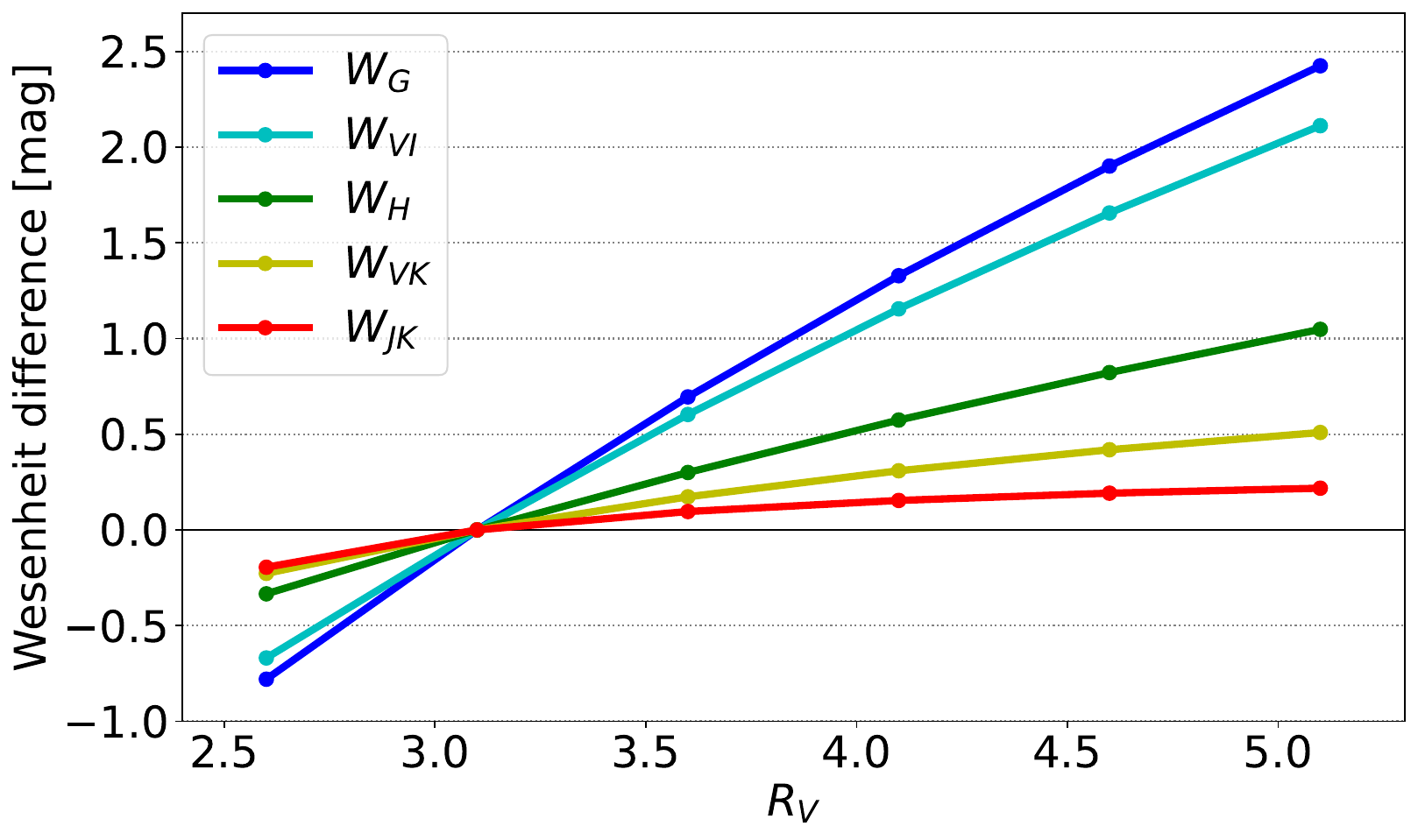}
\includegraphics[width=0.49\linewidth]{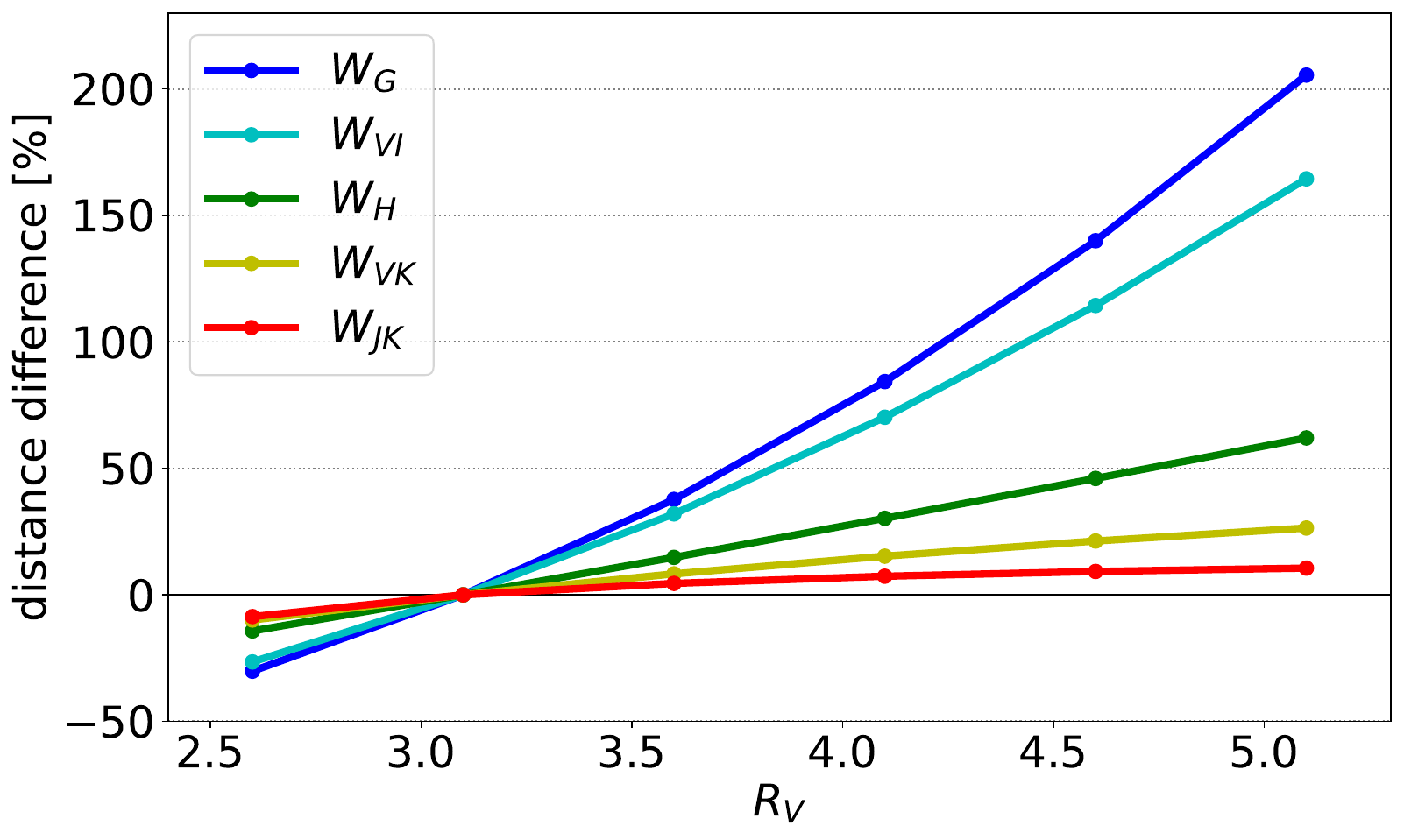}
\caption{The effect of $R_V$ on the Wesenheit index (left, Eqn.~\ref{eqn:wdiff}) and distance (right, Eqn.~\ref{eqn:ddiff}), relative to values at $R_V=3.1$. Each color represents
a different Wesenheit function $W$, defined in Table 1, according to the legend. Points are at $R_V$ values of 2.6, 3.1, 3.6, 4.1, 4.6, 5.1.}
\label{fig:magdistdiff}
\end{figure*}

\subsection{The impact of \rv\ on the Wesenheit magnitude and distance}
\label{sec:magdist}

Following the results of \cite{Zhang2025}, we adopt the mean value of $R_V= 3.1 \pm 0.22$. Then $R_V=2.6$ and $R_V=3.6$, which represent a typical \rv\ range in the Milky Way, are a good approximation of the $\pm2\sigma$ range around the mean of the integrated \rv\footnote{\cite{Zhang2025} provide maps for both differential, and integrated along a line-of-sight to a specific distance, $R_{55}$, from which the integrated \rv\ (Eq. \ref{eqn:rv}) can be derived.}
(see their Figure S8). In that case, the relative difference in the $R$ parameter within $R_V= 3.1 \pm 0.22$, estimated from Figure~\ref{fig:r_vs_rv}, is roughly 8.5\%, 10\%, 12\%, 21\%, and 21.5\% for $R_G$, $R_{VI}$, $R_{JK}$, $R_H$, and $R_{VK}$, respectively.

However, the final value of the Wesenheit magnitude depends not only on its $R$ coefficient, but also on the color of the star (Eqn.~\ref{eqn:wesenheit}). 
Therefore, to estimate how the Wesehneit magnitude of a given star would change with the change of \rv, we have to account for both factors. To do so, we calculate Wesenheit magnitudes at different values of \rv\ (estimating the adequate $R_{\lambda_1,\lambda_2}^{R_V}$ coefficients based on relations plotted in Figure~\ref{fig:r_vs_rv}), and at $R_V=3.1$:
\begin{align*}
\begin{split}
    W_{\lambda_1,\lambda_2} &= m_{\lambda_2} - R_{\lambda_1,\lambda_2}^{R_V} \times (m_{\lambda_1} - m_{\lambda_2})\\
    W_{\lambda_1,\lambda_2}^{3.1} &= m_{\lambda_2} - R_{\lambda_1,\lambda_2}^{3.1} \times (m_{\lambda_1} - m_{\lambda_2})
\end{split}
\end{align*}
and then their difference:
\begin{equation}
    W_{\lambda_1,\lambda_2} - W_{\lambda_1,\lambda_2}^{3.1} = (R_{\lambda_1,\lambda_2}^{3.1} - R_{\lambda_1,\lambda_2}^{R_V}) \times (m_{\lambda_1} - m_{\lambda_2})
    \label{eqn:wdiff}
\end{equation}

For the color term, we use mean colors and standard deviations of Milky Way Cepheids, which in the considered passband combinations are:
$(V-I) = 2.32 \pm 0.95$ mag, $(G_{BP}-G_{RP}) = 2.64 \pm 1.12$ mag, $(V-K) = 5.22 \pm 2.14$ mag, and $(J-K) = 1.08 \pm 0.47$ mag. We use these numbers together with $R_{\lambda_1,\lambda_2}$ coefficients for a given \rv\ value to plot the change in the Wesenheit index 
as a function of \rv\ (Eqn.~\ref{eqn:wdiff}), in the left panel of Figure~\ref{fig:magdistdiff}, where colors represent different Wesenheit functions. 
The right panel shows the resulting change in distance, relative to the distance calculated assuming $R_V=3.1$, as:
\begin{equation}
    \Delta d/ d = 10^{0.2(W_{\lambda_1,\lambda_2} - W_{\lambda_1,\lambda_2}^{3.1})}-1
    \label{eqn:ddiff}
\end{equation}
Since we calculate the change in $W$ and distance with respect to $R_V=3.1$, all lines cross at zero for $R_V=3.1$.

The largest differences are observed for the Gaia-based Wesenheit index, $W_G$, where a change within $R_V= 3.1 \pm 0.22$ results in $\pm0.37$~mag difference in $W_G$, which translates to $\pm17$\% change in distance. When considering the typically observed range of $(2.6-3.6)$, the discrepancies are more pronounced, e.g., for the \rv\ change between $3.1$ and $3.6$, the difference is $0.7$~mag in $W_G$, which translates to $38$\% distance change.
In the case of the $W_H$ index, the change within $R_V= 3.1 \pm 0.22$ is $\pm0.16$~mag in $W$ and $\pm7.5$\% in distance, while the change between $3.1$ and $3.6$ is $0.30$~mag in $W$ and $15$\% in distance. The smallest effect of variable \rv\ is observed for the near-infrared $W_{JK}$ index ($0.1$~mag and $4.5$\% in distance between $R_V= 3.1$ and $3.6$), which is not surprising, as the extinction is much smaller at infrared wavelengths than in the optical.

\section{Conclusions}

The analysis presented in this work demonstrates that the long-standing assumption of a universal extinction curve, expressed as a fixed value of \rv, has an underestimated negative impact on the Wesenheit magnitudes and PW-based distance estimates.
Using recent measurements of spatially varying \rv, we show that these variations introduce significant systematics in Cepheid Wesenheit magnitudes and distances.

Milky Way classical Cepheids are especially affected because they lie in the Galactic disk, where extinction is both high and highly variable. As a result, even moderate changes in \rv\ produce noticeable shifts in the $R$ coefficients used in optical Wesenheit relations. For Gaia-based Wesenheit magnitudes, $W_G$, a shift from $R_V = 3.1$ to $R_V = 3.6$ leads to changes of $\approx0.7$ mag in $W_G$, corresponding to distance differences of almost $40$\%. These shifts are large enough to explain the strong discrepancies found by \cite{Skowron2025} between Milky Way Cepheid distances derived from Gaia photometry and those obtained from mid-infrared PL relations combined with extinction maps. Near-infrared Wesenheit indices remain the least affected and therefore offer the most reliable distances.

PW relations constructed with different Wesenheit indices are also widely used for distance measurements in other galaxies, most notably in the Magellanic Clouds. As shown by \cite{Zhang2025}, the mean \rv\ in the SMC is lower than in the Milky Way, while in the LMC, the range of \rv\ is much wider.
Because the choice of $R$ directly affects the definition of the Wesenheit function, adopting an \rv\ appropriate for one galaxy but applying it to another will change the slope and zero point of the PW relation, and therefore alter the inferred distances. This would also affect the calibration of the extragalactic distance scale and, in consequence, influence measurements of the Hubble constant. For example, \cite{Riess2016} showed that $R$ variations arising from changing between the reddening laws of \cite{Fitzpatrick1999} and \cite{Cardelli1989} result in $H_0$ changing by $0.1-0.15$ km s$^{-1}$ Mpc$^{-1}$ when $W_H$ is used, and by $-2.15-3.8$ km s$^{-1}$ Mpc$^{-1}$ when $W_{VI}$ is used, demonstrating that near-infrared data remains the preferred choice for $H_0$ studies, in the absence of individual $R$ values.

The main conclusion of this work is that optical Wesenheit indices of Milky Way Cepheids are not "reddening-free", as the assumption of a universal extinction curve is not satisfied. Consequently, optical Wesenheit magnitudes should not be used to estimate distances to Milky Way Cepheids, and distance determinations should instead rely on near- or mid-infrared data, which are far less sensitive to variations in the extinction curve (e.g. \citealt{Skowron2025,Wang2025}).
However, a positive consequence is that the colors of highly reddened Cepheids contain information of \rv\ on large scales, as will be demonstrated in \citet[][in prep.]{Drimmel2025}. 
It is also important to note, that although our discussion focuses on classical Cepheids, the same considerations are valid for all classes of pulsating stars, such as RR~Lyrae or long-period variables, for which PW relations are employed to derive distances.

Finally, one could also ask about the continued use of Wesenheit functions in an era when obtaining high-quality multi-band photometry is no longer as big of a problem, as it was four decades ago, when the Wesenheit index was introduced. Multi-band approaches that simultaneously fit an extinction curve to apparent distance moduli across several wavelengths, introduced by \cite{Freedman1988} for $BVRI$ Cepheid photometry in IC~1613, make use of all available information and have been shown to provide reliable estimates of both distance and reddening for many galaxies.
This method was later extended by \cite{Rich2014} to a star-by-star basis. They calculated individual distance moduli to each Cepheid in NGC~6822 in seven bands ($VIJHK_S$, $3.6$ and $4.5 \mu m$), and fitted a reddening law (in distance-modulus versus inverse-wavelength space) for each star to determine the true distance modulus (corresponding to zero inverse wavelength). The Wesenheit index is nothing more than such a fit, but restricted to only two, or three, passbands. It is therefore straightforward that, when multi-band data are available, the Wesenheit function becomes unnecessarily restrictive.

Another important aspect of deriving Cepheid distances is taking into account wavelength-dependent effects such as metallicity, which has been shown to affect the PL and PW relations (e.g. \citealt{Bhuyan2026} and references therein). At the same time, metallicity is expected to correlate with dust properties, thus affecting the extinction curve parameters. Therefore, fully accounting for both the effects of variable \rv\ and metallicity, would require a complex covariant approach. While this is beyond the scope of this Letter, it represents an important issue that should be addressed in future studies.

\section{Software and third party data repository citations} \label{sec:cite}

\software{
 \texttt{scipy} \citep{2020SciPy-NMeth},
\texttt{numpy} \citep{harris2020array},
\texttt{matplotlib} \citep{Hunter:2007},
Jupyter notebooks \citep{Kluyver2016jupyter},
Pyphot \citep{Fouesneau_pyphot_2025},
}

\begin{acknowledgments}
The authors thank the Referee for insightful and constructive comments, which inspired a broader perspective on the subject of this Letter.
DMS acknowledges support from the European Union (ERC, LSP-MIST, 101040160). 
SK acknowledges support from the European Union's Horizon 2020 research and innovation program under the GaiaUnlimited project (grant agreement No 101004110).
Views and opinions expressed are, however, those of the authors only and do not necessarily reflect those of the European Union or the European Research Council. Neither the European Union nor the granting authority can be held responsible for them.
RD \& EP are supported in part by the Italian Space Agency (ASI) through contract 2025-10-HH.0 to the National Institute for Astrophysics (INAF).
This work made use of the Overleaf platform and the NASA Astrophysics Data System.
\end{acknowledgments}

\bibliography{paper}{}
\bibliographystyle{aasjournalv7}

\end{document}